\documentclass[%
manuscript,
superscriptaddress,
amsmath,amssymb,
aip,
apl,
floatfix,
]{revtex4-2}

\setcitestyle{super}
\usepackage{graphicx}
\usepackage{booktabs}
\usepackage{dcolumn}
\usepackage{bm}
\usepackage{makecell}
\usepackage{float}
\usepackage[T1]{fontenc}
\usepackage{xcolor}
\usepackage{textcomp,gensymb} 
\usepackage{soul}
\usepackage{rotating}
\usepackage{chngcntr}
\counterwithin*{figure}{section}
\usepackage{natbib}

\renewcommand{\figurename}{{\bf Figure}}
\renewcommand{\thefigure}{{\bf \arabic{figure}}}

\renewcommand{\tablename}{{\bf Table}}
\renewcommand{\thetable}{{\bf \arabic{table}}}

\usepackage{hyperref} 
\hypersetup{
    colorlinks=true,
    linkcolor=blue,
    filecolor=magenta,      
    urlcolor=cyan,
}
\newcounter{para}
\newcommand\N{\par\refstepcounter{para}}                                         

\renewcommand{\bibliography}[1]{} 
\begin{document}

\title{Laboratory evidence for proton energization by collisionless shock surfing}

\author{W. Yao}
\affiliation{LULI - CNRS, CEA, UPMC Univ Paris 06 : Sorbonne Universit\'e, Ecole Polytechnique, Institut Polytechnique de Paris - F-91128 Palaiseau cedex, France}
\affiliation{Sorbonne Universit\'e, Observatoire de Paris, Universit\'e PSL, CNRS, LERMA, F-75005, Paris, France}

\author{A. Fazzini}
\affiliation{LULI - CNRS, CEA, UPMC Univ Paris 06 : Sorbonne Universit\'e, Ecole Polytechnique, Institut Polytechnique de Paris - F-91128 Palaiseau cedex, France}

\author{S. N. Chen}
\affiliation{ELI-NP, "Horia Hulubei" National Institute for Physics and Nuclear Engineering, 30 Reactorului Street, RO-077125, Bucharest-Magurele, Romania}

\author{K. Burdonov}
\affiliation{LULI - CNRS, CEA, UPMC Univ Paris 06 : Sorbonne Universit\'e, Ecole Polytechnique, Institut Polytechnique de Paris - F-91128 Palaiseau cedex, France}
\affiliation{Sorbonne Universit\'e, Observatoire de Paris, Universit\'e PSL, CNRS, LERMA, F-75005, Paris, France}
\affiliation{IAP, Russian Academy of Sciences, 603155, Nizhny Novgorod, Russia}

\author{P. Antici}
\affiliation{INRS-EMT, 1650 boul, Lionel-Boulet, Varennes, QC, J3X 1S2, Canada}

\author{J. B\'eard}
\affiliation{LNCMI, UPR 3228, CNRS-UGA-UPS-INSA, Toulouse 31400, France}

\author{S. Bola\~{n}os}
\affiliation{LULI - CNRS, CEA, UPMC Univ Paris 06 : Sorbonne Universit\'e, Ecole Polytechnique, Institut Polytechnique de Paris - F-91128 Palaiseau cedex, France}

\author{A. Ciardi}
\affiliation{Sorbonne Universit\'e, Observatoire de Paris, Universit\'e PSL, CNRS, LERMA, F-75005, Paris, France}

\author{R. Diab}
\affiliation{LULI - CNRS, CEA, UPMC Univ Paris 06 : Sorbonne Universit\'e, Ecole Polytechnique, Institut Polytechnique de Paris - F-91128 Palaiseau cedex, France}

\author{E.D. Filippov}
\affiliation{JIHT, Russian Academy of Sciences, 125412, Moscow, Russia}
\affiliation{IAP, Russian Academy of Sciences, 603155, Nizhny Novgorod, Russia}

\author{S. Kisyov}
\affiliation{ELI-NP, "Horia Hulubei" National Institute for Physics and Nuclear Engineering, 30 Reactorului Street, RO-077125, Bucharest-Magurele, Romania}

\author{V. Lelasseux}
\affiliation{LULI - CNRS, CEA, UPMC Univ Paris 06 : Sorbonne Universit\'e, Ecole Polytechnique, Institut Polytechnique de Paris - F-91128 Palaiseau cedex, France}

\author{M. Miceli}
\affiliation{Universit\`a degli Studi di Palermo, Dipartimento di Fisica e Chimica E. Segr\`e, Piazza del Parlamento 1, 90134 Palermo, Italy}
\affiliation{INAF–Osservatorio Astronomico di Palermo, Palermo, Italy}

\author{Q. Moreno}
\affiliation{University of Bordeaux, Centre Lasers Intenses et Applications, CNRS, CEA, UMR 5107, F-33405 Talence, France}
\affiliation{ELI-Beamlines, Institute of Physics, Czech Academy of Sciences, 5 Kvetna 835, 25241 Dolni Brezany, Czech Republic}

\author{V. Nastasa}
\affiliation{ELI-NP, "Horia Hulubei" National Institute for Physics and Nuclear Engineering, 30 Reactorului Street, RO-077125, Bucharest-Magurele, Romania}

\author{S.~Orlando}
\affiliation{INAF–Osservatorio Astronomico di Palermo, Palermo, Italy}

\author{S. Pikuz}
\affiliation{JIHT, Russian Academy of Sciences, 125412, Moscow, Russia}
\affiliation{NRNU MEPhI, 115409, Moscow, Russia}

\author{D. C. Popescu}
\affiliation{ELI-NP, "Horia Hulubei" National Institute for Physics and Nuclear Engineering, 30 Reactorului Street, RO-077125, Bucharest-Magurele, Romania}

\author{G. Revet}
\affiliation{LULI - CNRS, CEA, UPMC Univ Paris 06 : Sorbonne Universit\'e, Ecole Polytechnique, Institut Polytechnique de Paris - F-91128 Palaiseau cedex, France}

\author{X. Ribeyre}
\affiliation{University of Bordeaux, Centre Lasers Intenses et Applications, CNRS, CEA, UMR 5107, F-33405 Talence, France}

\author{E. d'Humi\`eres}
\affiliation{University of Bordeaux, Centre Lasers Intenses et Applications, CNRS, CEA, UMR 5107, F-33405 Talence, France}

\author{J. Fuchs}
\affiliation{LULI - CNRS, CEA, UPMC Univ Paris 06 : Sorbonne Universit\'e, Ecole Polytechnique, Institut Polytechnique de Paris - F-91128 Palaiseau cedex, France}

\date{\today}

\maketitle

\textbf{Charged particles can be accelerated to high energies by collisionless shock waves in astrophysical environments, such as supernova remnants. By interacting with the magnetized ambient medium, these shocks can transfer energy to particles. Despite increasing efforts in the characterization of these shocks from satellite measurements at the Earth’s bow shock and powerful numerical simulations, the underlying acceleration mechanism or a combination thereof is still widely debated. Here, we show that astrophysically relevant super-critical quasi-perpendicular magnetized collisionless shocks can be produced and characterized in the laboratory. We observe characteristics of super-criticality in the shock profile as well as the energization of protons picked up from the ambient gas to hundreds of keV. Kinetic simulations modelling the laboratory experiment identified shock surfing as the proton acceleration mechanism. Our observations not only provide the direct evidence of early stage ion energization by collisionless shocks, but they also highlight the role this particular mechanism plays in energizing ambient ions to feed further stages of acceleration. Furthermore, our results open the door to future laboratory experiments investigating the possible transition to other mechanisms, when increasing the magnetic field strength, or the effect induced shock front ripples could have on acceleration processes.}

The acceleration of energetic charged particles by collisionless shock waves is an ubiquitous phenomenon in astrophysical environments, e.g. during the expansion of supernova remnants (SNRs) in the interstellar medium (ISM) \cite{2009Sci...325..719H}, during solar wind interaction with the Earth's magnetosphere \cite{turner}, or with the ISM (at the so-called termination shock) \cite{decker2008voyager2}. \textcolor{black}{In SNRs, there is a growing consensus that the acceleration is efficient at quasi-parallel shocks \cite{2014ApJ...783...91C,reynoso2013radio}, while in interplanetary shocks, the quasi-perpendicular scenario (i.e. the magnetic field is perpendicular to the shock normal, or the on-axis shock propagation direction) is invoked \cite{burrows2010pickup,zank2009microstructure,chalov2016acceleration}.} \textcolor{black}{The quasi-perpendicular shocks that produce particle acceleration are qualified as super-critical; they have a specific characteristic such that in addition to dissipation by thermalisation and entropy, energy is dissipated also by reflecting the upstream plasma. According to \cite{coroniti1970dissipation,edmiston1984parametric}, the threshold for the super-critical regime of the quasi-perpendicular shock is defined as: $M_{ms} = v_{sh} / \sqrt{v_A^2 + C_s^2} \gtrsim 2.7$  (where $M_{ms}$ is the magnetosonic Mach number, $v_{sh}$, $v_A$ and $C_s$ are the shock, Alfv\'{e}nic and sound velocity, respectively). } 

Three basic ion acceleration mechanisms are commonly considered to be induced by such shocks\cite{blandford1987particle,balogh2013physics,Vlahosth2016microphysics}: diffusive shock acceleration (DSA), shock surfing acceleration (SSA), and shock drift acceleration (SDA). \textcolor{black}{The first proceeds with particles gaining energy by scattering off magnetic perturbations present in the shock upstream and downstream media}, whereas, in SSA and SDA, the particles gyro-rotate (Larmor motion) in close proximity with the shock and gain energy through the induced electric field associated with the shock. 

These last two processes are mostly differentiated by how and where the particles are trapped around the shock front and the ratio of the ion Larmor radius vs. the shock width \textcolor{black}{(large for SSA and small for SDA) \cite{shapiro2003shock,yang2012impact}. And here in our case we expect that SSA dominates over SDA, as will be detailed below.}

DSA, which is commonly invoked for high-energy particle acceleration in SNRs, is thought to require quite energetic particles to be effective \cite{zank1996interstellar,lee1996pickup}, which raises the so-called ``injection problem'' of their generation \cite{lembege2004selected}. Providing those pre-accelerated seed particles is precisely thought to be accomplished by SSA or SDA, which are evoked to accelerate particle at low energies, e.g. in our solar system \cite{burrows2010pickup}.

Due to the small sampling of such phenomena even close to Earth, the complexity of the structuring of such shocks \cite{lee1996pickup,2015ApJ...798L..28C}, and the related difficulty in modelling them realistically, the question of the effectiveness and relative importance of SDA and SSA \cite{yang2012impact} is still largely debated in the literature.

We will first show that laboratory experiments can be performed to generate and characterize globally mildly super-critical, quasi-perpendicular magnetized collisionless shocks. The shock shown in Fig.~\ref{fig:optical_probe} is typically produced by using a laser-driven piston to send an expanding plasma into an ambient (a cloud of hydrogen) secondary plasma \cite{schaeffer2019direct} in an externally controlled, \textcolor{black}{homogeneous and highly reproducible} magnetic field (see Methods). The high-strength applied magnetic field \cite{albertazzi2013production} we use is key in order to ensure the collisionless nature of the induced shock. The key parameters of the laboratory created shock are summarized in Table~\ref{tab:solar_para}, which shows that they compare favorably with the parameters of the Earth's bow shock \cite{ellison1990particle,turner}, the solar wind termination shock \cite{richardson2008cool,burlaga2008magnetic,decker2008voyager2}, and \textcolor{black}{of four different} non-relativistic SNRs interacting with dense molecular clouds \textcolor{black}{(see Extended Data Table 1 detailing the considered objects)}.

A snapshot of the integrated plasma electron density was obtained by optical probing at 4 ns after the laser irradiation of the target and is shown in Fig.~\ref{fig:optical_probe}c and d in the two perpendicular planes containing the main expansion axis. The laser comes from the right side and the piston source target is located at the left (at $x=0$).We can clearly see both the piston front and the shock front (indicated by the orange and green arrows, respectively). A lineout of the plasma density is shown in Fig.~\ref{fig:optical_probe}e, where the piston and shock fronts are well identified by the abrupt density changes. The piston front is steepened by the compression induced by the magnetic field \cite{khiar2019laser} (see also Extended Data Fig. 5). In contrast, when the B-field is switched off, only a smooth plasma expansion into the \textcolor{black}{ambient gas} (blue dashed line) can be seen. In the case when the magnetic field is applied, another clear signature of the magnetized shock, as observed by satellites crossing the Earth's bow shock \cite{giagkiozis2017validation}, is the noticeable feature of a ``foot'' in the density profile, located in the shock upstream (US). It is due to the cyclic evolution of the plasma: the plasma in the foot is picked up to form the shock front, while the front itself is also periodically dismantled by the Larmor \textcolor{black}{motion of the ions}.  The observed foot width is of the order 0.5-1 mm, which compares favourably with the expected foot width being twice the ion inertial length \cite{baraka} (which is here $\approx 0.23$ mm), and with the width observed in our simulations shown below.

Fig.~\ref{fig:optical_probe}f shows the shock front position evolution and the corresponding velocity deduced from it, which shows the very fast decrease of the shock velocity over the first few ns. Before 2.6 ns, \textcolor{black}{the shock front velocity is around $V_0 = 1500$ km/s, corresponding to an ion-ion collisional mean-free-path $\lambda_{mfp} = V_0 \tau_i \approx 10$ mm, with the ion collisional time $\tau_i \approx 6$ ns} -- both are larger than the interaction \textcolor{black}{spatial and temporal scales}, indicating that the shock is collisionless. \textcolor{black}{Note also that for such velocity, the Larmor radius of the ions in the shock is around \textcolor{black}{0.8} mm, i.e. \textcolor{black}{larger than} the shock width, which, although it is too small to be well resolved by our interferometer, is well below 0.2 mm, suggesting favourable conditions for SSA to be at play.} However, after 4-5 ns, the shock velocity decreases rapidly to about 500 km/s, thus becoming sub-critical and the foot of the shock becomes less distinguishable (see also in Extended Data Fig. 1). \textcolor{black}{Later we will demonstrate, with the help of kinetic simulations, that the proton acceleration happens within the first 2-3 ns of the shock evolution, i.e. when the shock is super-critical, with a front velocity above 1000 km/s.}


The plasma temperature was measured at a fixed location at different instants in time (see Methods), allowing to characterize the temperature increase in the shock as it swept through the probed volume, as shown in Fig. ~\ref{fig:TS_diag}. Before the shock front, \textcolor{black}{the electron temperature $T_e$ is around 70 eV \textcolor{black}{(consistent with the heating induced by the Thomson scattering laser probe, see Extended Data Fig. 2)} and ion temperature $T_i$ is about 20 eV.} While behind the shock front, $T_e$ is almost doubled (see Fig.~\ref{fig:TS_diag}a), $T_i$ is increased dramatically to about 200 eV, and $T_i$ becomes larger than $T_e$. All of the above results are typical signs of a shock wave. {\color{black}Again, the formation of the shock is only possible due to the applied external magnetic field. In its absence, as shown in Extended Data Fig. 3, we witness no ion temperature increase in the same region.} Extended Data Fig. 4 shows the electron density increase in the shock compared to that of the \textcolor{black}{ambient gas}.

Another important aspect of our experiment is the observation of non-thermal protons when the piston interacts with the magnetized ambient gas. The recorded spectra, shown in Fig.~\ref{fig:Spectrum} (red dots), \textcolor{black}{clearly show the presence of non-thermal proton energization when the external magnetic field is applied, i.e. with a spectral slope significantly larger than that of the thermal proton spectrum of 200 eV, which is represented by the red dash-dot line.} \textcolor{black}{The cutoff energy reaches to about $80$ keV, close to the Hillas limit \cite{hillas1984origin, drury2012origin} (an estimate of the maximum energy that can be gained in the acceleration region, which is around 100 keV with the velocity of 1500 km/s and the acceleration length around 3-4 mm in the first 2 ns, as shown in Fig.~\ref{fig:optical_probe}f)}. {\color{black}We stress that} without the external B-field or {\color{black}in the absence of} ambient gas, no signal is recorded in the ion spectrometer (hidden under the experimental noise baseline, as indicated by the cyan dashed line in Fig.~\ref{fig:Spectrum}).


{\color{black}Fig.~\ref{fig:optical_probe}b shows the result of a 3D magneto-hydrodynamic (MHD) simulation (performed using the FLASH code, see Methods) of the experiment. We observe that it reproduces globally the macroscopic expansion of the piston in the magnetized ambient gas and the shock formation (see also Extended Data Fig. 6). However, no foot can be observed. What is more, in the MHD simulation, the shock velocity is quite steady and does not show the strong and fast energy damping experienced by the shock from the experiment. Both facts point to a non-hydrodynamic origin of the foot and of the energy loss experienced by the shock in its initial phase. This is why, in the following, we resort to kinetic simulations with a Particle-In-Cell (PIC) code (the fully kinetic code SMILEI, see Methods). \textcolor{black}{The PIC simulation focuses on the dynamics of the shock front (already detached from the piston) and of its interaction with the ambient gas, using directly the shock parameters measured in the experiment, and not that of the MHD simulation}. As the shock changes from super- to sub-critical in its evolution, we have performed simulations in two cases (see Fig.~\ref{fig:Spectrum}), i.e. with two different velocities representative of the two phases, i.e. 1500 and 500 km/s respectively} to {\color{black}unveil the micro-physics} responsible for the observed non-thermal proton acceleration.

Note that in order to directly compare with the experimental spectra, the ion specie in our simulation is proton with its real mass ($m_p / m_e = 1836$).
The PIC simulation results, for the high-velocity case, are summarized in Fig.~\ref{fig:PIC_result}, which identifies clearly the underlying proton acceleration mechanism, matching the laboratory proton spectrum (see Fig.~\ref{fig:Spectrum}), to be SSA \cite{katsouleas1983unlimited}. 

\N
Fig.~\ref{fig:PIC_result}a illustrates the overall evolution of the early stage of the high-velocity shock. Shown is the proton density in the reference frame of the contact discontinuity (CD), where we can clearly see the density pileups in the forward direction, indicating the shock formation (and periodic reformation \cite{balogh2013physics}). \textcolor{black}{To elucidate the proton acceleration mechanism, a random sample of protons ($10^4$ out of $10^7$) are followed in the simulation. More than 2\% of those end up with energies $> 40$ keV, which will constitute the high-energy end of the spectrum shown in Fig.~\ref{fig:Spectrum}. They share similar trajectories and two representative ones (P1 and P2) are plotted in Fig.~\ref{fig:PIC_result}a.} Following these trajectories, we can see that they are first picked up by the forward shock at the shock front, and then they gain energy while ``surfing'' along (or confined around) the shock front. Besides, while surfing along the shock front, P1 gets trapped and reflected repeatedly, with a small energy perturbation, as is shown in Fig.~\ref{fig:PIC_result}b, all of which is typical of SSA.

\N
Typical structures of the super-critical quasi-perpendicular collisionless shock \cite{balogh2013physics} can also be seen in Fig.~\ref{fig:PIC_result}c where we plot the lineout of density and electromagnetic (EM) fields around the shock front ($0.4 < x < 1.2$ mm in the reference frame of the CD), when the shock is fully formed ($t=1.5$ ns).

\N
The longitudinal electric field ($E_x$) is seen to peak right at the ramp, providing the electrostatic cross-shock potential to trap and reflect the protons (with a velocity lower than that of the shock). In the corresponding $x-v_x$ phase space in Fig.~\ref{fig:PIC_result}d, we can clearly see that, indeed, it is at the position of $E_x$ that protons get reflected (see also Extended Data Fig. 7 for more time frames of this phase space, as well as for those corresponding to the simulation performed at low-velocity). This rules out the possibility of SDA, where the ion reflection is caused by the downstream compressed B-field \cite{zank1996interstellar}, being dominant. At last, as shown in Fig.~\ref{fig:PIC_result}e, the main contribution of the proton energy gain is due to $v_y$ via the inductive electric field $E_y = v_x B_z$, which is again in accordance with the SSA mechanism \cite{matsukiyo2011microstructure}.

\N
The proton spectrum at $t=2.6$ ns produced by the PIC simulated high-velocity shock, and which is shown in Fig.~\ref{fig:Spectrum} (black solid line), is in remarkable agreement with the experimental observation. {\color{black}As in the experiment, no proton energization is found in the simulations performed without magnetic field or ambient medium. We note also that} for the low-velocity shock (with $v = 500$ km/s), the spectrum (green dashed line) is far below the experimental noise baseline, indicating that the protons are indeed accelerated at the first $2-3$ ns, when the shock is in the super-critical regime. \textcolor{black}{The energy spectra of other simulation cases can be found in Extended Data Fig. 8, proving the robustness of our results.}

Hence, a remarkable outcome of our analysis is that, \textcolor{black}{for the parameters at play in our experiment and} at the early stage of the shock formation and development, SSA can be considered as the sole mechanism in picking up thermal ions and accelerating them to hundred keV-scale energies. SSA appears to produce sufficiently energetic protons for further acceleration by DSA, as for example at the Earth's bow shock, where the threshold energy for DSA to become effective is in the range of $\sim (50-100)$ keV/nucleon \cite{balogh2013physics}. Since we are limited in time in exploring the dynamics of the protons interacting with the super-critical shock, we can only speculate that SDA might appear at a later stage, when the reflected ions acquire enough energy to cross the shock front. 

We also note that usually detailed considerations of shock rippling and structuration are evoked in a possible competition between SSA and SDA in the solar wind \cite{chalov2016acceleration,yang2012impact}, but that these were not required here in our analysis where we simulate an idealized flat shock front. Although we know that in the experiment, there is likely small structuring developing at the shock front (induced by instabilities \cite{khiar2019laser}, but too small at this early stage to be resolved by our optical probing), these are obviously not required in the modelling to reproduce the experimentally observed energization.  

\N
Aside from the solar wind, another interesting case of a shock similar to that investigated here is that of supernova remnants (SNRs) interacting with dense molecular clouds, e.g. the class of Mixed-Morphology SNRs \cite{rho}. A large fraction of these SNRs show indications of low energy (MeV) cosmic rays (CRs) interacting with the cloud material and ionising it \cite{Nobukawa2019,2019MNRAS.484.2684N,Okon2020}. These mildly relativistic particles are typically explained as CRs accelerated in the past at the SNR shock front that escaped the remnant and reached the cloud\cite{phan2020}. However, our results show that in-situ generation of low energy CRs ($\sim$ MeV) could be at play, and should also be taken into account \cite{Nobukawa2019}. The in-situ acceleration would be most likely generated by the low-velocity, mildly super-critical (see Table~\ref{tab:solar_para}) SNR shock interacting with the dense cloud; a scenario which is supported by our findings\textcolor{black}{: since our analysis of the experiment shows that SSA is most likely behind the observed proton energization, and since the plasma parameters at play in the experiment are similar to those of the objects detailed in Extended Data Table 1, we suggest that SSA is similarly effective in these objects}. 

\N
In conclusion, our experiment provides strong evidence for the generation of super-critical quasi-perpendicular magnetized collisionless shocks in the laboratory. More importantly, non-thermal proton spectra are observed; in our kinetic simulations, they are recognized to be produced by SSA alone. \textcolor{black}{Such efforts for proton acceleration, together with those for electrons \cite{rigby2018electron, li2019collisionless,fiuza_electron_2020}, will certainly shed new light on the ``injection problem'' in astrophysically-related collisionless shocks \cite{lebedev2019exploring}}.

The platform we used can be tuned in the future to monitor the transition to DSA, which should be favored by varying the magnetic field orientation, using even higher-strength magnetic field \cite{fujioka} or higher-velocity jets driven by short-pulse lasers as pistons \cite{kar}. Another direction will be to test quantitatively the effect of intentionally rippling the shock front by seeding the piston plasma with modulations \cite{cole}.




\section*{Acknowledgments}
The authors would like to thank the teams of the LULI (France) and JLF laser (USA) facilities for their expert support, as well the Dresden High Magnetic Field Laboratory at Helmholtz-Zentrum Dresden-Rossendorf for the development of the pulsed power generator used at LULI. We thank the Smilei dev-team for technical support. We also thank Ph. Savoini (Sorbonne U., France), L. Gremillet, C. Ruyer, P. Loiseau (CEA-France), and M. Manuel (General Atomics, USA) for discussions. W.Y. would like to thank R. Li (SZTU, China) for discussions. This work was supported by funding from the European Research Council (ERC) under the European Unions Horizon 2020 research and innovation program (Grant Agreement No. 787539, JF).
The computational resources of this work were supported by the National Sciences and Engineering Research Council of Canada (NSERC) and Compute Canada (Job: pve-323-ac, PA).
Part of the experimental system is covered by a patent (1000183285, 2013, INPI-France). 
The FLASH software used was developed, in part, by the DOE NNSA ASC- and the DOE Office of Science ASCR-supported Flash Center for Computational Science at the University of Chicago.
The research leading to these results is supported by Extreme Light Infrastructure Nuclear Physics (ELI-NP) Phase II, a project co-financed by the Romanian Government and European Union through the European Regional Development Fund (SNC, SK, VN, DCP), and by the project ELI-RO-2020-23 funded by IFA (Romania, JF) 
JIHT RAS team members are supported by The Ministry of Science and Higher Education of the Russian Federation (Agreement with Joint Institute for High Temperatures RAS No 075-15-2020-785, EDF, SP). 
The reported study was funded by the Russian Foundation for Basic Research, project No. 19-32-60008 (EDF, SP).

\section*{Author Contributions Statement}

J.F. and S.N.C. conceived the project. A.F., S.N.C., K.B., J.B., S.B., S.K., V.L., V.N., S.P., D.C.P., G.R. and J.F. performed the experiments. A.F., E.D.F., S.N.C., K.B., R.D., S.P. and J.F. analyzed the data. X.R. performed and analyzed the FLASH simulations, while W.Y. and A.F. performed and analyzed the SMILEI simulations, both with discussions with P.A., A.C., Q.M., X.R., E.d.H. and J.F. W.Y., S.N.C., and J.F. wrote the bulk of the paper, with major contributions from K.B., M.M. and S.O. All authors commented and revised the paper.

\textbf{\textit{Materials and correspondence}}
Correspondence and material requests should be addressed to julien.fuchs@polytechnique.edu

\section*{Competing Interests Statement}
The authors declare no competing interests.

\newpage 

\section*{Figure legends/captions}

\begin{figure}[htp]
    \centering
    \caption{\textbf{Configuration and characterization of the laboratory super-critical shock.}
    (a) Cartoon of the setup of the experiment performed at JLF/Titan to characterize the shock. \textcolor{black}{The ambient gas is indicated by the grey color around the plasma plume, which is indicated by the orange hemisphere.} 
    (b) Three-dimensional (3D) rendering of the ablated piston \textcolor{black}{electron number density ($N_e$ in the unit of cm$^{-3}$)} interacting with the magnetized ambient gas at 4 ns from the FLASH simulation (see Methods). The orange arrow represents the piston front, while the green one the shock front.
    (c) and (d) Experimental density measurement (integrated along the line of sight) 4 ns after the laser irradiation of the solid target (at $t=0$) recorded in the two different and complementary xy and xz planes in order to characterize in three-dimensions the overall plasma. Each image corresponds to a different laser shot. More images at different times are shown in Extended Data Fig. 1. \textcolor{black}{The magnetic field direction is indicated in both panels with B (OUT) and B (VERT), respectively, which is both along the z-direction. The black solid lines are the lineouts for panel (e).}
    (e) Lineouts obtained along the black solid lines from panel (c) and (d) (indicated with the red and orange solid line, respectively), comparing to the unmagnetized case (the blue dashed line with $B = 0$).
    (f) Evolution of the shock front position along the x-axis, and the corresponding velocity.  Each point corresponds to the average of the shock front of all relevant shots. The error bars on the x position represent not only the maximum extent of the variation of the shock front position as observed in relevant shots, but they also integrate the uncertainty of the initial target surface ($x=0$) as well as the width of the shock front. The error bars on the velocity correspond to the propagation of the errors on the position. The small relative error attests to the reproducibility of the overall experimental phenomenon.
    }
    \label{fig:optical_probe}
\end{figure}

\begin{figure}[htp]
    \centering
    \caption{\textbf{Laboratory characterization of electron and ion temperature in the shock.} The measurements are performed at LULI2000, using collective Thomson scattering (TS) on the electron and ion waves in the plasma, in a fixed volume 4.3 mm away from the solid target surface, and with $B = 20$ T applied (see Methods). Time 0, at which a jump is identified, here corresponds to the time at which the shock is sweeping through the location of the measurement. Each data point corresponds to a shot.
    Panel (a) illustrates the local electron temperature inferred from both the measurements on the electron and ion waves. Panel (b) corresponds to the local ion temperature inferred from the measurement on the ion waves. The vertical error bars reflect the variations of the parameters when fitting the data with a theoretical fit, while still fitting well the data (see examples in Extended Data Fig. 3 and 4). The horizontal bar reflects the duration (3 ns) of the laser beam used to perform the measurement.}
    \label{fig:TS_diag}
\end{figure}

\begin{figure}[htp]
    \centering
    \caption{\textbf{Evidence for the energization of protons picked up from the \textcolor{black}{ambient medium}.} Proton energy spectra of both the experiment (red dots, averaged over five shots, as measured at LULI2000) and of two PIC simulations (\textcolor{black}{the black solid line for the high-velocity case with $v=1500$ km/s and the yellow dashed line} for the low-velocity one with $v=500$ km/s, both are measured at $t=2.6$ ns in the simulations). 
    \textcolor{black}{The horizontal axis ($E_p$) represents the kinetic energy of the protons, while the vertical axis represents the number of protons per bin of energy ($dN/dE$), divided by the solid angle ($d\Omega$) subtended by the entrance pinhole of the spectrometer (in the case of the experimental spectrum).} The red dash-dot line is the thermal proton spectrum of 200 eV.
    The blue error bars correspond to one sigma deviation from the average (shown by the red dots), the noise level on the diagnostic is materialized by the cyan \textcolor{black}{dotted} line. 
    Note that the absolute scale in proton numbers applies only to the experimental spectrum\textcolor{black}{, which is represented by the red dots}; the simulated spectrum is arbitrarily scaled to the experiment one.}
    \label{fig:Spectrum}
\end{figure}

\begin{figure}[htp]
    \centering
    \caption{\textbf{Dynamics of the shock surfing proton energization from PIC simulations.} 
    (a) \textcolor{black}{Trajectories of two protons (P1 and P2, energized from the ambient gas to $> 40$ keV at t=2.6 ns, representing 108 out of 5000 tracked particles with the ``surfing/reflection/gyromotion'' trajectories}) in $x-t$ diagram, overlaid on the proton density map \textcolor{black}{in the reference frame of the contact discontinuity.} \textcolor{black}{The colormap on the top represents the proton energy E in the unit of keV; while the colormap on the right represents the proton number density $n_i$ in the unit of $10^{18}$ cm$^{-3}$.}
    (b) Zoom of the black dashed rectangle region in (a), showing the ``surfing \& reflection'' of P1 along the forward shock. 
    (c) Lineout of density \textcolor{black}{($n_i$ in black dashed line)} and electromagnetic fields \textcolor{black}{($B_z$ in red solid line, $E_y$ in blue dashed line, and $E_x$ in purple solid line)} (\textcolor{black}{normalized by each of their maximum value respectively}) at the red dashed line in (a) ($t=1.5$ ns, $0.4 < x < 1.2$ mm). \textcolor{black}{Note that we also show them separately with their units in SI in Extended Data Fig. 9.}
    (d) The corresponding proton $x-v_x$ phase space diagram where the colorbar represents the normalized particle number N in logarithm scale.
    (e) The $v_x-v_y$ diagram of P1 and P2. The grey shaded area corresponds to the ``surfing \& reflection" stage in (b).}
    \label{fig:PIC_result}
\end{figure}



\begin{table}[htp]
\centering
\begin{tabular}{p{0.3\textwidth}|p{0.1\textwidth}<{\centering}|p{0.15\textwidth}<{\centering}}
\hline
\textbf{}                                       & \textcolor{black}{$\boldsymbol{M_{ms}}$} & $\boldsymbol{\lambda_{mfp}/r_{L,i}}$ \\ \hline
\textbf{\makecell{Our experiment}}              & \textcolor{black}{3.1}                & \textcolor{black}{12.2}                      \\ \hline
\textbf{\makecell{Bow Shock}}                   & \textcolor{black}{2.8}                & $1.2\times10^8$                      \\ \hline
\textbf{\makecell{Term. Shock}}                 & \textcolor{black}{4.9}                & $7.4\times10^7$                      \\ \hline
\textbf{\makecell{Mixed-morphology \\ SNR}}     & \textcolor{black}{$\ge 3.2$}           & $\ge 2.0\times 10^2$                \\ \hline
\end{tabular}%
\caption{\textbf{Parameters of the laboratory shocks and three natural ones.} The three natural events are: the Earth's bow shock, the solar wind termination shock, and low-velocity interactions of mixed-morphology SNRs with dense molecular clouds. The listed parameters for each plasma are: 
\textcolor{black}{the Magnetosonic Mach number ($M_{ms}$, see text),}
and the ratio of the collisional mean free path ($\lambda_{mfp}$) over the ion Larmor radius ($r_{L,i}$). The first shows that all shocks are super-critical, the second shows that all shocks are collisionless, i.e. electromagnetic forces dominate over collisions since the collision mean free path is always much \textcolor{black}{larger} than the ion Larmor radius. Details on the parameters and how they are derived are given in Extended Data Table 1.}
\label{tab:solar_para}
\end{table}

\newpage 

\section*{Methods}
\subsection*{Experimental setup}

\N
The experiment was performed at the JLF/Titan (LLNL, USA) and LULI2000 (France) laser facilities with similar laser conditions but using complementary diagnostics, which was mostly linked with the availability of different auxiliary laser beams at each facility. 

\N
In the experiment at JLF/Titan, we used a high-power laser pulse (1 $\mu$m wavelength, 1 ns duration, 70 J energy, $1.6 \times 10^{13}$ W/cm$^2$ intensity on target) to irradiate a solid target (Teflon, CF$_2$). We used this material in order to exploit the x-ray emission from ionized F ions in the expanding \textcolor{black}{piston} plasma in order to diagnose, through x-ray spectroscopy (as detailed below), the properties of the laser-ablated plasma. The laser irradiation induced the expansion of a hot plasma (the piston) that propagates along the x-axis, as is shown in Fig.1a of the main text. Prior to the shot, a large volume gas jet is pulsed from a nozzle, so that the whole scene is homogeneously (i.e. over larger scales than that shown in Fig.1 of the main text and in Extended Data Fig. 1) embedded in an H$_2$ gas of low density ($10^{18}$ cm$^{-3}$). Further, the whole assembly is embedded in a strong magnetic field (20 T) oriented along the z-axis, which is created by a Helmholtz coil system \cite{higginson2019laboratory, albertazzi2013production}. \textcolor{black}{The created magnetic field in our experiment is spatially uniform within 5\% at the scale of shock acceleration (i.e. within 5 mm distance from the initial target surface) \cite{higginson2017detailed}. 
} \textcolor{black}{Also note that the created magnetic field in our experiment typically varies by less than 1\% from shot-to-shot \cite{albertazzi2013production}}. The collisionless shock is formed as a consequence of the plasma piston propagating in the magnetized ambient gas. The plasma electron density is recorded by optically probing the plasma (using a mJ, 1 ps auxiliary laser pulse) and using an interferometry setup, as detailed in \cite{higginson2017detailed}. It allows to measure electron plasma densities in the range $10^{17}$ cm$^{-3}$ to a few $10^{19}$ cm$^{-3}$ (limited by the refraction of the optical probe beam in the steep density gradients close to the initial solid target surface). Since the system is three-dimensional and does not present an axis of symmetry, due to the presence of the magnetic field along the z-axis that breaks the symmetry of the system, we probe the plasma along two different axes (x and y). This allows us to measure plasma density maps (integrated along the line of sight) in the xy and xz planes, as shown in Fig.1. A complete time sequence of such maps is shown in Extended Data Fig. 1.

\N
The experiment performed at LULI2000 had similar laser conditions (1 $\mu$m wavelength, 1 ns, 100 J, but we kept the same on-target intensity, i.e. $1.6 \times 10^{13}$ W/cm$^2$, by adjusting the laser spot size on target). At LULI2000, a second high-energy auxiliary (0.5265 $\mu$m wavelength, 1 ns, 15 J, focused over $\sim$ 40 $\mu$m along the z-axis and propagated throughout the plasma) was available, allowing us to perform Thomson scattering (TS) off the electron and ion waves in the plasma. To accommodate the laser beams inside the narrow space within the magnetic field generation coil, the target had to be tilted by 45\degree \ around the z-axis and lifted up (along y). This led to the target surface to be outside the optical probing field of view of the probe beam, but allowed to see the similar development of the piston and shock wave as at JLF/Titan, and then to perform TS. TS was used in a mode where the plasma was sampled in a collective mode \cite{froula2011plasma}, the collection of the scattered light being performed at 90\degree \ (along the y-axis) from the incident direction of the laser probe (the z-axis). With TS, we could access spatially and temporally resolved measurements of the plasma density and temperatures (electron and ion) in the upstream (US), as well as in the downstream (DS) region. \textcolor{black}{Note that the Thomson scattering laser probe induces some heating in the hydrogen ambient gas. The electron temperature resulting from such heating through inverse Bremsstrahlung absorption can be analytically estimated (see caption of Extended Data Fig. 2 for details). For this, we estimate the in-plasma intensity of the Thomson laser probe as 1.5 $\times 10^{13}$ W$\cdot$cm$^{-2}$, based on the fact that we see it being defocused to at least 200 microns diameter. The result is shown in Extended Data Fig. 2. It suggests that the upper limit of the TS laser probe-induced heating is around 60 eV, i.e. consistent with the values we actually measure prior to the passing of the shock front in the observation location. Note that such value is also significantly smaller than the level of temperatures we observe induced by the shock.}

\N
The light scattered off the ion (TSi) and electron (TSe) waves in the plasma was analyzed by means of two different spectrometers, set to different dispersions (3.1 mm/nm for TSi and $7.5 \times 10^{-2}$ mm/nm for TSe), and which were coupled to two streak-cameras (Hamamatsu for TSe, and TitanLabs for TSi, both equipped with S-20 photocathode to be sensitive in the visible part of the spectrum, and both with typical 30 ps temporal resolution), allowing us to analyze the evolution of the TS emission in time.

\N
The central openings of both streak-cameras and spectrometers were imaging at the same location in the plasma (located 4.3 mm away from the solid target surface) within the magnetic field coil, in order to ensure that the value of the electron density obtained from the TSe analysis corresponds to the same region of the plasma that was observed in the corresponding TSi spectrum.

\N
Thin strips of black filters were positioned at the entrance slits of both streak cameras to block the Rayleigh scattered light at the wavelength of the probing laser, for both TSi and TSe.
The scattering volumes sampled by the instruments were: 120 $\mu$m along the x-axis, 120 $\mu$m along the y-axis 40 $\mu$m along the z-axis for TSi; 150 $\mu$m along the x-axis, 100 $\mu$m along the y-axis, 40 $\mu$m along the z-axis for TSe.
\N
The analysis of the TS was performed by comparison of the experimental images (recorded by the streak cameras) with the theoretical equation of the scattered spectrum for coherent TS in unmagnetized and non-collisional plasmas, with the instrumental function taken into an account. Examples are shown in Extended Data Fig. 3 and Extended Data Fig. 4. Such analysis allowed us to yield the electron density, as well as the electron and the ion temperatures of the plasma in the probed volume\cite{froula2007quenching,froula2011plasma}. 

\N
The x-ray emission was measured by a Focusing Spectrometer with high Spatial Resolution (FSSR) \cite{Faenov:1994} at both laser facilities. It was based on a spherically-bent mica crystal, with $2d=19.9149$~\r{A} and $R=150$~mm, to detect H- and He-like spectral lines of fluoride ions \textcolor{black}{(of the piston)} in the range 750--1000~eV. The spectrometer was installed in the direction transverse to the plasma propagation, having a spatial resolution ($\sim$ 100~$\mu$m) over more than 10~mm along the plasma expansion axis. Fluorescent detector Fujifilm Image Plate TR covered by an aluminized Mylar filter against emission in the visible range was used as a detector. The spectral resolution ${\lambda}/{\Delta}{\lambda}$ better than 1000 was achieved. The signal is time-integrated.

\N
Similarly to the electron density data measured by optical interferometry (shown in the main text), the x-ray spectrometer demonstrates a spatial profiles of spectral lines confirming the spatial compression of the piston when the magnetic field in applied, compared to the case of the non-magnetized ambient gas jet (see Ly${\alpha}$ emissivities in Extended Data Fig. 5). The presence of He-like spectral lines of Fluorine allowed us to measure the parameters of the \textcolor{black}{piston}, using the quasi-stationary method described in details in \cite{Ryazantsev:2016}. \textcolor{black}{This is complementary to the Thomson scattering measurements, which characterize the shock in the hydrogen plasma.} 

\N
Last, an ion spectrometer, having a permanent magnet of 0.5 T and equipped with a pinhole, was deployed along the axis of the magnetic field (the z-axis) in an alternate mode to perform TS. The ions were detected using absolutely calibrated imaging plates as detectors \cite{manvcic2008absolute}. Having the spectrometer collection axis aligned with that of the magnetic field allows to measure the ions energized out of the plasma  \cite{higginson2019laboratory}, which otherwise could not be recorded, as they would be deflected away by the 20 T large-scale B-field. We used filters in order to eliminate the possibility that the signal observed in the dispersion plane of the spectrometer was originating from heavy ions others than protons from the ambient gas. No signal above the noise level (marked by the  cyan dashed line in Fig. 3 of the main text) was recorded in the spectrometer in the absence of the ambient hydrogen gas, neither when no magnetic field was applied.

\subsection*{Magneto-hydrodynamics simulations performed using the FLASH code}

\N
The experiment was modeled with the 3D Magneto-Hydrodynamic (MHD) code FLASH~\cite{fryxell2000flash}, to study the dynamics of the plasma plume expansion and shock formation in the ambient gas jet medium and in the presence of the strong magnetic field. We performed the simulations in 3D geometry, using 3 temperatures (2 for the plasma, and one for the radiation) with EOS and radiative transport, in the frame of ideal MHD and including the Biermann battery mechanism of magnetic field self-generation in the plasmas \cite{haines1986magnetic}. 

\N
The initial configuration of the simulations, modelled after that of the experiment, is the following: the laser beam is normal to a Teflon target foil and has an on-target intensity of $10^{13}$ W/cm$^2$. The generated plasma plume expands in the hydrogen gas-jet having an uniform density of $10^{18}$ cm$ ^{-3}$. Moreover, the plasma plume expands in the uniform external magnetic field of 20 Tesla (aligned along the z-axis, as in the experiment).

\textcolor{black}{Since the shock condition is dominated by kinetic effects, there are discrepancies between the hydrodynamic simulations and the experiments in the shock temperatures, especially for the ions. This is why we have resorted to using PIC simulations, the initial conditions of which are taken from the experimental measurements. Nevertheless, we can still observe that the FLASH simulations reproduce well the dynamics of the piston that induces the shock.}

\N
Extended Data Fig. 6 shows the magnetic field configuration of the expanding plasma. In particular, Extended Data Fig. 6a shows the plasma piston which expels the magnetic field, creating a bubble that is void of the magnetic field \cite{khiar2019laser}. This piston launches a shock inside the ambient gas, with the magnetic field upstream of the shock being compressed because of magnetic flux conservation. This is clearly shown in Extended Data Fig. 6b,  where the magnetic field lines are plotted. Because of the three dimensional nature of the plasma flow and of the magnetic pressure, the piston and the shocked ambient plasma are more compressed by the magnetic pressure along the y-axis than along the z-axis.

\N
\textcolor{black}{From the density map at 10 ns (after the laser irradiation) of the FLASH simulation, we can still clearly see the continuously expanding plasma plume (and the compressed Teflon target) because once the laser energy is deposited on the target surface, the heat wave and the shock propagation inside the target (due to collisional effects) can last for much longer time (tens of ns) \cite{atzeni2004physics}. Such heat wave drives continuous ablation after the laser irradiation.}
 
\N

\subsection*{Particle-In-Cell simulations performed using the SMILEI code}

\N
The microscopic particle dynamics is modelled with the kinetic PIC code SMILEI \cite{derouillat2018smilei}. 
Since the scale of the shock front interaction with the ambient medium is much smaller across the shock (a few hundreds of microns) than along the shock (a few mm), i.e. the interaction is quasi one-dimensional (1D), we use the 1D3V version of the code, and initialize our simulation box to be $L_x = 2048 d_e = 11$ mm, with the spatial resolution $d_x = 0.2 d_e = 1.1 \ \mu$m, where $d_e = c / \omega_{pe} = 5.3 \ \mu$m is the electron inertial length, and $\omega_{pe} = (n_{e0} q_e^2 / m_e / \epsilon_0)^{1/2} = 5.6 \times 10^{13}$ s$^{-1}$ is the electron plasma frequency. Here, c is the speed of light, $n_{e0} = 1.0 \times 10^{18}$ cm$^{-3}$ is the electron number density of the ambient plasma, and $m_e$, $q_e$ and $\epsilon_0$ are the electron mass, elementary charge, and the permittivity of free space, respectively. 
\textcolor{black}{With the above simulation units, the ion mean-free-path is $\lambda_{mfp} \approx 1800 d_e$, which is larger than the interaction scale. The Debye length is small compared to the spatial resolution, i.e. $\lambda_{De} = (\epsilon_0 kT_{e0}/n_{e0} q_e^2)^{1/2} \approx 0.01d_e = 0.05 d_x$, with the initial low temperature of the ambient plasma ($T_{e0} = 50$ eV). However, with a series of cases using different temperatures, it is proved that the shock dynamics is robust.}
The magnetic field is homogeneously applied in the z-direction with $B_z = 20$ T ($\omega_{ce} / \omega_{pe} = 0.06$, where $\omega_{ce} = q_e B / m_e$), \textcolor{black}{leading to a Larmor radius $r_{Li} = 0.8$ mm and a proton gyro-period $\tau_{Li} = 3$ ns} 
(for the high-velocity case of 1500 km/s \textcolor{black}{in the upstream)}.

\N
For each cell, we put 1024 particles for each specie. The simulation lasts for $1.5 \times 10^5 \omega_{pe}^{-1} \sim 2.5$ ns. \textcolor{black}{Fields are absorbed, while particles are removed at the boundaries,}
and enough room is left between the boundary and the shock, so that the boundary conditions do not affect the concerned physics.
The ambient plasma lies in the right half of the simulation box, while the left half is for the shocked plasma, flowing towards the right with an initial velocity of $v_1 = 1500$ km/s (high-velocity case) or $v_1 = 500$ km/s (low-velocity case). \textcolor{black}{The shock width is initialized as one ion inertial length $d_i \sim 200 \ \mu$m in between the ambient plasma and the shocked plasma.}

\N
Other parameters of the shocked plasma flow are: 
the electron number density is $n_{e1} = 2n_{e0} = 2.0 \times 10^{18}$ cm$^{-3}$, and the temperature is $T_{e1} = 100$ eV and $T_{i1} = 200$ eV, all inferred from the TS characterization (see Fig. 2 of the main text).  Note that in our simulation, the ion specie is proton with its real mass ($m_p / m_e = 1836$). We have also tested a series of 1D and 2D simulations with  varying resolutions ($0.1 \sim 0.5d_e$), particle-per-cell numbers ($512 \sim 2048$), and ion-to-mass ratio ($400 \sim 1836$), all reaching similar shock behavior as detailed in the main text, which proves the robustness of the observed behavior of the quasi-perpendicular, super-critical magnetized collisionless shock investigated here.

\N
The phase space evolution of the shock interaction with the ambient is shown in Extended Data Fig. 7 for the considered two cases having different flow velocities.
The first row corresponds to dynamics of the high-velocity case ($v_1 = 1500$ km/s), from the formation of both the forward (FS) and reverse  (RS) shocks at $t = 0.5$ ns in (a), to the FS reformation at $t = 1.5$ ns in (b), when protons within the downstream region start to gyrate within the compressed B-field. At last, while the FS shock remains picking up protons from the ambient gas, the downstream RS becomes highly nonlinear. 

\N
Comparing the first row of Extended Data Fig. 7 with the second row, which corresponds to the low-velocity case ($v_1 = 500$ km/s), it is clear that the reflected protons are much less in the latter case due to the low \textcolor{black}{Magnetosonic Mach number $M_{ms} = v/\sqrt{v_A^2 + C_s^2} \lesssim 1.0 < M_c = 2.7$}.
Although there is also shock reformation at $t = 1.5$ ns in (e), at last in (f), the downstream region is just heated a little bit and the FS remains in its linear stage, where even the magnetosonic wave pattern can be seen traveling forward (because $M_{ms} \lesssim 1$). This sub-critical result of the low-velocity case is in accordance with the results of experiments performed at the Gekko XII laser facility \cite{shoji2016toward}.

\N
As shown in Extended Data Fig. 8a, the energy spectrum from our experimental data can be fitted by a power-law function $\propto E^p$ with index $p = -4.28$. This is quite smaller than the usual index of shock acceleration (e.g. for DSA, $p \sim -2.3$; while for SDA, $p \sim -1.5$ \cite{zank1996interstellar}). The reason is because the shock in the experiment is short lived. We can speculate that if the shock would have been longer lived (e.g. if the laser driving the plasma ablation would have been longer), the protons could have been accelerated to higher energy and the index could have reached higher values.
We confirm the above by doing a simulation that lasts for twice longer time (see the green simulated spectrum in Extended Data Fig. 8a), and it is clear that at $t=5.1$ ns, the energy spectrum gets flatten and the index reaches $p=-2.27$, which is getting closer to the index inferred from the analysis of astrophysical observations \cite{decker2005voyager}.
\N
As a check, we have verified that the shock and the overall dynamics that we are investigating is collisionless by verifying that we obtained almost identical energy spectra in simulations performed with and without ion-ion collisions (see Extended Data Fig. 8b).

In order to further verify the robustness of our mechanism, we have done a series of cases with different B-field strength, as is shown in Extended Data Fig. 8c. It shows that the acceleration efficiency is larger with higher strength of B-field, which is in accordance with the SSA.


\textcolor{black}{Since the density and EM fields are each normalized by their respective maximum value in Fig. 4c of the main paper (in order to compare their relative values in a single plot), here we also show them separately with their units in SI. This is shown in Extended Data Fig. 9a-e (to clearly demonstrate their values).} \textcolor{black}{In addition, we also have simulations that last for longer time (twice of the former simulation, i.e. around 5.3 ns, with twice the former simulation box size, i.e. around 22 mm). As is shown in Extended Data Fig. 9f, in this case, the protons have time to complete several gyro cycles, and we can observe that this does not change the physical picture we could infer from the shorter duration simulation, the one shown in the paper.}

\subsection*{Astrophysical relevance}

Table 1 of the main paper as well as the detailed version of it presented in the Extended Data Table 1 show that the shock in our experiment has a relevance with those of the Earth bow shock, of the solar wind termination shock, and of the low-velocity SNRs interacting with dense molecular clouds. 

To begin with, all systems are collisionless due to the fact that the collision mean free path ($\lambda_{mfp}$) is much smaller than the ion Larmor radius ($r_{L,i}$), which leads to the system being dominated by electromagnetic forces rather than collisions.
Additionally, all systems have \textcolor{black}{Magnetosonic Mach numbers over 2.7}, indicating that the plasma is qualified as super-critical \cite{balogh2013physics}, i.e. the plasma then is subject to additional dissipation in the form of ions extracting energy from the shock front from which they are reflected and accelerated.
Besides, since all systems have Reynolds numbers much larger than unity, we can ensure that all systems are dominated by turbulence.
Also, the magnetic Reynolds number is much larger than unity in all cases, indicating that for all systems magnetic advection dominates over magnetic diffusion. 
\textcolor{black}{In addition}, since we have the Peclet number much larger than unity in all cases, we can ensure that the flow itself is dominated by advection rather than diffusion.
\textcolor{black}{Last but not least, in the case of the low-velocity SNRs interacting with dense molecular clouds, e.g. W44, the ion Larmor radius is only in the order of $10^5$ cm (see Extended Data Table 1), whereas its shock width has been measured to be much larger, i.e. in the order of $10^{17} - 10^{18}$ cm \cite{cosentino2019interstellar}. This further supports that SSA can be more efficient than SDA in these cases.}

\bigskip
\textbf{\textit{Data availability}}
All data needed to evaluate the conclusions in the paper are present in the paper.
Experimental data and simulations are respectively archived on servers at LULI and LERMA laboratories and are available from the corresponding author upon reasonable request.

\bigskip
\textbf{\textit{Code availability}}
The code used to generate Fig.~\ref{fig:optical_probe}b is FLASH. The code used to generate part of Fig.~\ref{fig:Spectrum} and Fig.~\ref{fig:PIC_result} is SMILEI. Both codes are detailed in the Methods section.

\section*{Extended Data Figure legends/captions}

\renewcommand{\figurename}{{\bf Extended Data Figure}}
\renewcommand{\thefigure}{{\bf \arabic{figure}}}

\setcounter{figure}{0}

\begin{figure}[htp]
    \centering
    \includegraphics[width=1.0\textwidth]{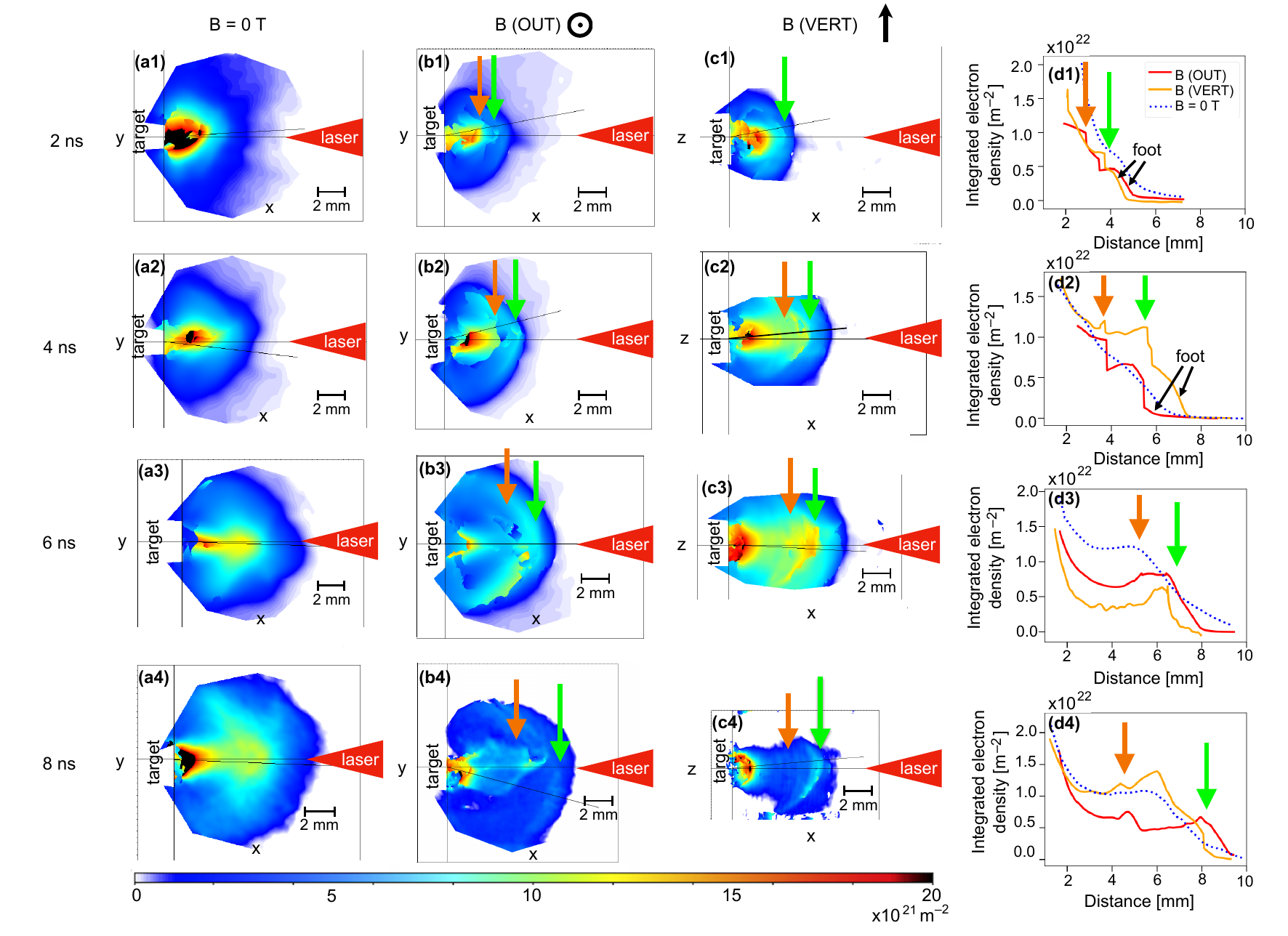}
    \caption{\textbf{Time sequence of experimental density measurement} (integrated along the line of sight) recorded in the two different and complementary xy and xz planes in order to characterize in three-dimensions the overall plasma. Each image corresponds to a different laser shot. \textcolor{black}{Specifically, the first column is for the case without the external magnetic field; the second and third column are for the case with the external magnetic field in the xy- and xz-plane, respectively.}
    The color scale shown at the bottom applies to all images. The corresponding lineouts along the thin dark lines shown in each image are shown on the \textcolor{black}{fourth column. From top to bottom, each row represents a different time, i.e., 2/4/6/8 ns.} \textcolor{black}{Magnetic field directions are indicated at the top of each column. Orange and green arrows indicate the piston and the shock front, respectively.}
    }
    \label{fig:intdens26-8}
\end{figure}

\begin{figure}[htp]
    \centering
    \includegraphics[width=0.5\textwidth]{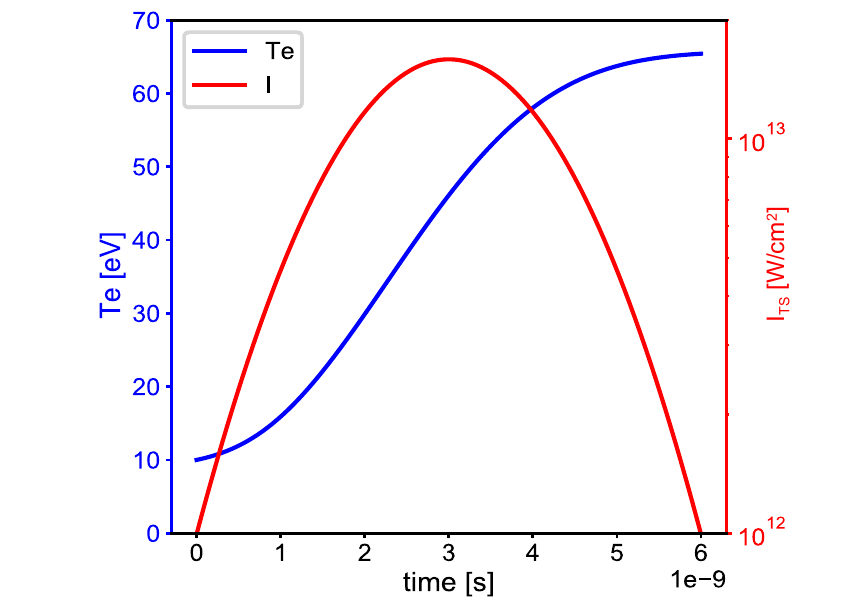}
    \caption{\textbf{Temporal evolution of the electron temperature} obtained from the solution of $1.5n_e d_t T_e = \nu_B I_0 e^{-t^2/\tau^2}$. Here, $n_e = 10^{18}$ cm$^{-3}$, $\nu_B$ is the inverse Bremsstrahlung absorption coefficient from the NRL formulary (p.58/Eq.32). 
    The laser energy is 15 J,  with duration $\tau = 3$ ns, defocused focal spot $d = 200  \mu$m and wavelength $\lambda_{l} = 528$ nm, leading to an maximum intensity of $I_0 = 1.5 \times 10^{13}$ W$\cdot$cm$^{-2}$, and the initial electron temperature $T_{e0} = 10$ eV (left axis, blue). The intensity evolution of the laser ($I_{TS}$) is superimposed as a red line (right axis).}
    \label{fig:TS_heating}
\end{figure}
    
\begin{figure}[htp]
    \centering
    \includegraphics[width=\textwidth]{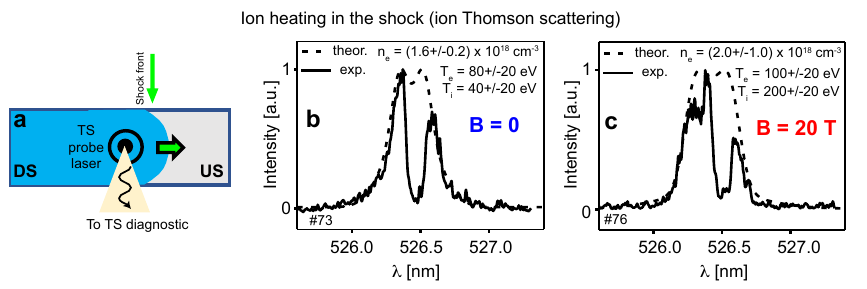}
    \caption{\textbf{Examples of plasma density and temperature measurement} of the shock \textcolor{black}{downstream (DS)}. (a) Schematic diagram of the collective Thomson scattering (TS) diagnostic deployed at LULI2000. The measurement is performed, in a fixed volume (see Methods) through which the shock is sweeping, 4.3 mm away from the solid target surface.
    Panels (b) and (c) show TS measurement on the ion waves in the plasma, allowing to retrieve the local electron and ion temperatures, as stated. Both measurements are performed 3 ns after the shock has passed, but (b) corresponds to the case without external B-field, while (c) corresponds to the case with $B = 20$ T applied. Solid lines -- experimental data profiles; dashed lines -- theoretical spectra. The stated uncertainties in the retrieved plasma parameters represent the possible variation of the parameters of the theoretical fit, while still fitting well the data, as well as the shot-to-shot variations observed in the same conditions.}
    \label{fig:TS_diag_ion_heat}
\end{figure}

\begin{figure}[htp]
    \centering
    \includegraphics[width=0.8\textwidth]{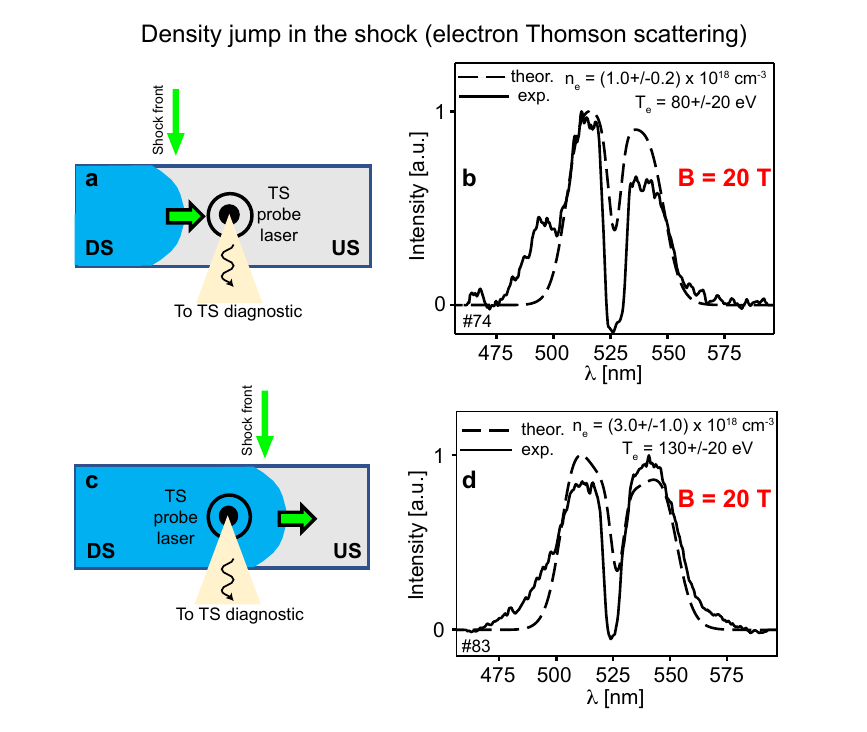}
    \caption{\textbf{Examples of plasma density and temperature measurement} \textcolor{black}{upstream (US) and downstream (DS) of the shock.} (a) and (c) Schematic diagram of the collective Thomson scattering (TS) diagnostic deployed at LULI2000. The measurement is performed, in a fixed volume (see Methods) through which the shock is sweeping, 4.3 mm away from the solid target surface. Panel (b) illustrates a TS measurement on the electron waves in the plasma, allowing the retrieve the local electron density and temperature, as stated. The measurement is performed 3 ns before the shock sweeps through (as illustrated in (a)). Panel (d) corresponds to the measurement performed at the same location, 1 ns after the shock has passed (as illustrated in (c)). 
    Solid lines -- experimental data profiles; dashed lines -- theoretical spectra. The stated uncertainties in the retrieved plasma parameters represent the possible variation of the parameters of the theoretical fit, while still fitting well the data, as well as the shot-to-shot variations observed in the same conditions.}
    \label{fig:TS_diag_den_jump}
\end{figure}

\begin{figure}[htp]
    \centering
    \includegraphics[width=0.5\textwidth]{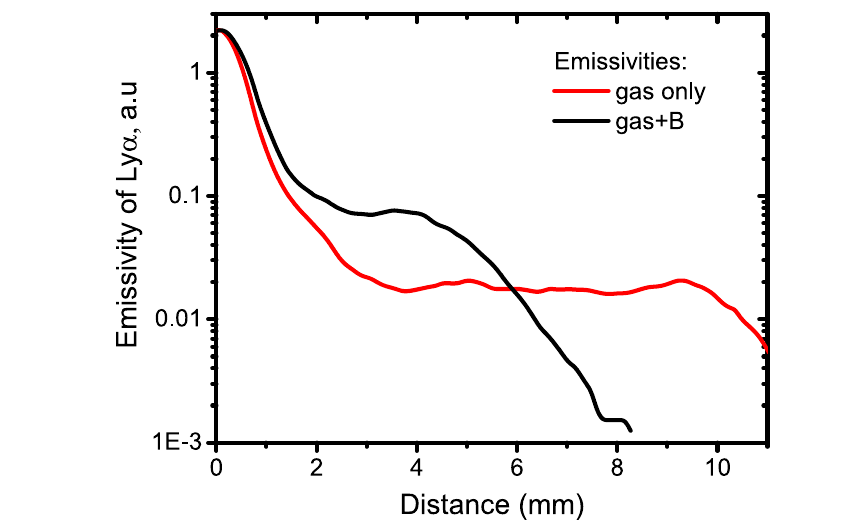}
    \caption{\textbf{X-ray emissivity of Ly$\alpha$ line measured by the FSSR.}
    The red curve represents the \textcolor{black}{piston} plasma emission in the range $0$--$10$~mm in the case when the unmagnetized \textcolor{black}{piston} expands in the ambient gas. The black curve represents the same but when additionally, the transverse external magnetic field was applied. 
    }
    \label{fig:FSSR}
\end{figure} 

\begin{figure}[htp]
    \centering
    \includegraphics[width=0.8\textwidth]{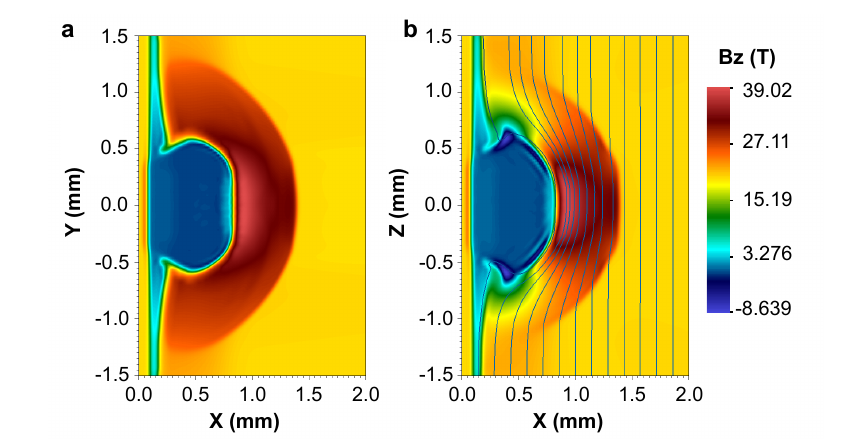}
    \caption{\textbf{Maps of the magnetic field in the MHD simulations}, 2 ns after the start of the plasma piston expansion inside the magnetized ambient medium. Shown are two-dimensional cross-section in the (a) xy and (b) xz planes, the latter with the corresponding magnetic field lines. The target is located on the left side of the box and the laser comes from the right side, as in the experiment. \textcolor{black}{The colormap represents the strength of $B_z$ in the unit of Tesla.}}
    \label{fig:Sim_B_field}
\end{figure}

\begin{figure}[htp]
    \centering
    \includegraphics[width=\textwidth]{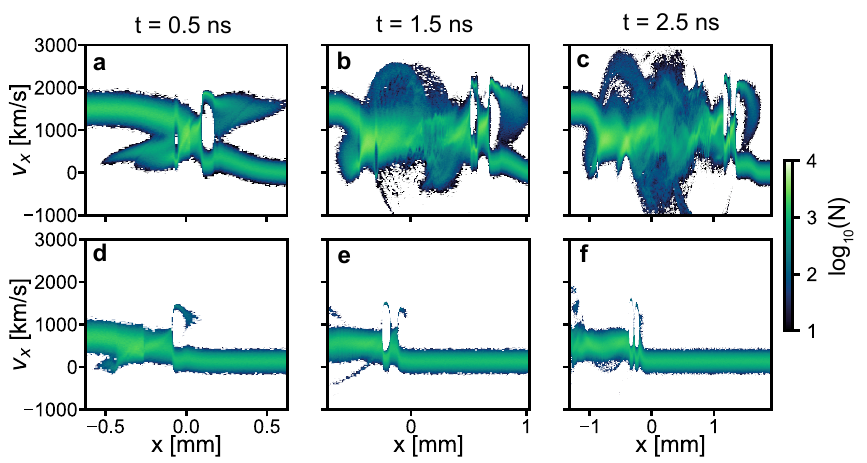}
    \caption{\textbf{Proton phase space evolution in the PIC simulations.} \textcolor{black}{The horizontal axis ($x$) is the proton position and the vertical axis ($v_x$) is the proton velocity along the x-direction.} The first row corresponds to the high-velocity case \textcolor{black}{with initial velocity of 1500 km/s}, while the second row is for the low-velocity one \textcolor{black}{with initial velocity of 500 km/s} (see text for more details), at (a) \& (d) 0.5 ns, (b) \& (e) 1.5 ns, and (c) \& (f) 2.5 ns. The colorbar represents the normalized particle number N in logarithm scale.}
    \label{fig:ps_evo}
\end{figure}

\begin{figure}[htp]
    \centering
    \includegraphics[width=\textwidth]{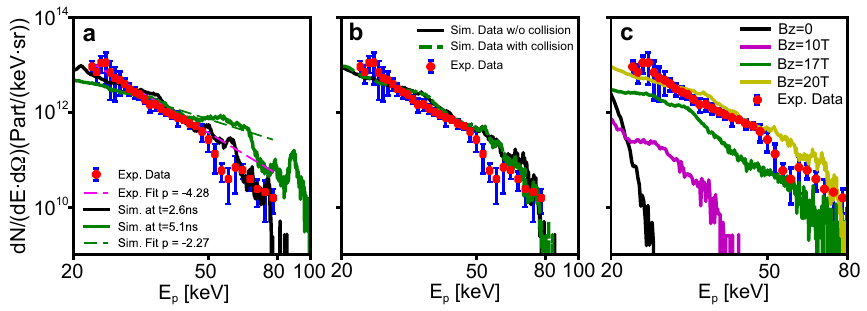}
    \caption{\textbf{Extended proton energy spectra.} (a) \textcolor{black}{Energy spectrum from the experiment represented by red dots (averaged over 5 shots) with blue error bars (correspond to one sigma deviation from the average), fitted by a power-law function (purple dashed line)}; together with PIC simulated spectra at $t=2.6$ ns (black solid line) and $t=5.1$ ns (green solid line). \textcolor{black}{The horizontal axis ($E_p$) represents the kinetic energy of the protons, while the vertical axis represents the number of protons per bin of energy ($dN/dE$), divided by the solid angle ($d\Omega$) subtended by the entrance pinhole of the spectrometer (in the case of the experimental spectrum).} Note that the absolute scale in proton numbers applies only to the experimental spectrum; the simulated spectra are adjusted to the experimental one. The latter is also fitted by a power-law function (green dashed line). (b) The same experimental data, as well as the simulations with and w/o collisions (green dashed line and black solid line, respectively). (c) Energy spectra of protons in cases of different B-field strength, \textcolor{black}{overlaid on the same experimental data}. The B-field strength is varied by varying the angle between the B-field direction (along z) and the on-axis shock propagation direction (along x) in the xz-plane.
    }
    \label{SI_fig:spectrum_fit}
\end{figure}

\begin{figure}[htp]
    \centering
    \includegraphics[width=\textwidth]{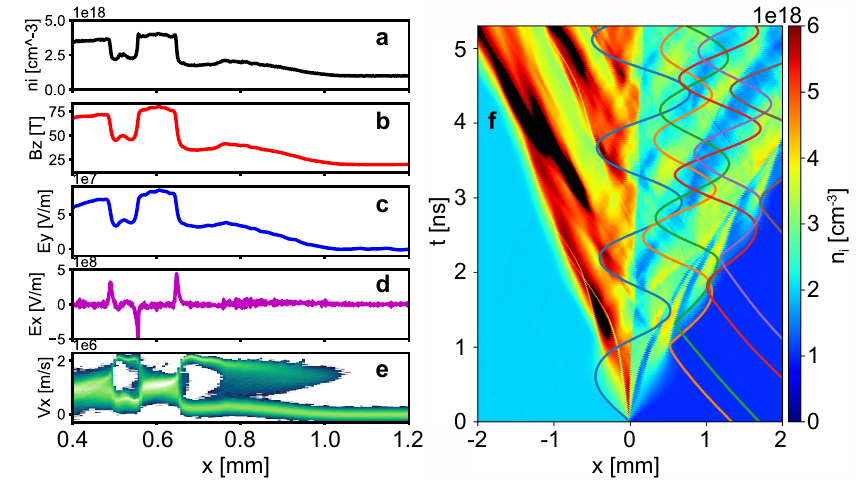}
    \caption{\textbf{Extended PIC simulation results.} (a)-(e) Lineouts of density ($n_i$) and EM fields ($B_z$, $E_y$, and $E_x$), as well as the corresponding phase-space distribution ($V_x$) at the same position and time as in Fig.4c in the main paper, i.e., from 0.4 mm to 1.2 mm at 1.5 ns, in SI units. (f) Trajectories of six randomly selected protons energized from the ambient gas in the $x-t$ diagram, overlaid on the proton density map in the contact discontinuity reference frame; all this for a simulation run over 5.31 ns, i.e., over longer time than the simulation shown in the paper.}
    \label{fig:units}
\end{figure}

\section*{Extended Data Tables}

\renewcommand{\tablename}{{\bf Extended Data Table}}
\renewcommand{\thetable}{{\bf \arabic{table}}}
\setcounter{table}{0}

\begin{sidewaystable}[htp]
\setlength{\tabcolsep}{8pt}
\renewcommand{\arraystretch}{0.5}
\scriptsize
\begin{tabular}{@{}c|c|c|c|c|c|c|c|c@{}}
\toprule
\shortstack{Parameters} & Our Results & \makecell{Earth's\\Bow Shock} & \makecell{Solar Wind\\Term. Shock} & \makecell{Mixed morpho. \\ SNR W28} & \makecell{Mixed morpho. \\ SNR Kes. 78 \cite{miceli2017xmm}} & \makecell{Mixed morpho. \\ SNR W44} & \makecell{Mixed morpho. \\ SNR IC443 \cite{greco2018discovery}} 
\\ \midrule
Flow Velocity $V$ [cm/s] & $\mathbf{1.5\times 10^8}$ & $\mathbf{5.0\times 10^7}$\cite{slavin1981solar} & $\mathbf{3.5\times 10^7}$\cite{richardson2008cool} & $\mathbf{(2.0-3.0)\times 10^6}$\cite{Velaz2002,vaupre2014} & $\mathbf{1.0\times 10^7}$\cite{boumis2009discovery} & $\mathbf{5.0\times 10^6}$ \cite{frail1998oh} & $\mathbf{6.0\times 10^6}$\cite{cesarsky1999isocam,reach2019supernova} 
\\
Magnetic Field $B$ [G] & $\mathbf{2.0\times 10^5}$ & $\mathbf{2.5\times 10^{-4}}$\cite{stasiewicz2020quasi} & $\mathbf{1.0\times 10^{-6}}$\cite{burlaga2008magnetic} & $\mathbf{(0.4-7.0)\times 10^{-4}}$\cite{hoffman2005oh, brogan2007oh, li2010gamma, abdo2010fermi,phan2020} & $\mathbf{1.5\times 10^{-3}}$\cite{koralesky1998shock} & $\mathbf{4.0\times 10^{-4}}$\cite{claussen1997polarization} & $\mathbf{3.0\times 10^{-4}}$\cite{reach2019supernova} 
\\
\makecell{Electron Temperature \\ $T_e$ [eV]} & $\mathbf{1.0\times 10^2}$ & $\mathbf{5.0}$\cite{slavin1981solar} & $\mathbf{1.0}$ & $\mathbf{0.1-1.0}$\cite{phan2020,Okon2018} & $\mathbf{1.0}$\cite{boumis2009discovery} & $\mathbf{0.1-1.0}$\cite{frail1998oh} & $\mathbf{1.0}$\cite{cesarsky1999isocam,reach2019supernova} 
\\
Ion Temperature $T_i$ [eV] & $\mathbf{2.0\times 10^2}$ & $\mathbf{1.5\times 10^{1}}$\cite{slavin1981solar} & & & & & \\
\makecell{Electron Number Density \\ $n_e$ [$cm^{-3}$]} & $\mathbf{1.0\times 10^{18}}$ & $\mathbf{1.0\times 10^{1}}$\cite{balogh2013physics} & $\mathbf{1.0\times 10^{-3}}$ & $\mathbf{1.0 \times 10^{3}}$\cite{vaupre2014,phan2020} & $\mathbf{1.0\times 10^{5}}$\cite{lockett1999oh} & $\mathbf{1.0\times 10^{4}}$\cite{frail1998oh} & $\mathbf{1.0\times 10^{4}}$\cite{cesarsky1999isocam,reach2019supernova} \\
\makecell{Characteristic Length Scale \\ $L_0$ [cm]} & $\mathbf{1.0\times 10^{-1}}$ & $\mathbf{1.0\times 10^{7}}$\cite{bale2003density,liebert2018statistical} & $\mathbf{6.0 \times 10^{8}}$\cite{burlaga2008magnetic} & $\mathbf{6.0\times 10^{19}}$\cite{Velaz2002} & $\mathbf{6.0\times 10^{19}}$ & $\mathbf{6.0\times 10^{19}}$ & $\mathbf{6.0\times 10^{19}}$ \\ \midrule
Sound Velocity $C_s$ [cm/s] & $2.2\times 10^{7}$ & $6.0\times 10^{6}$ & $ 1.8 \times 10^{6}$ & {$(0.6-1.8)\times 10^{6}$} & $1.8\times 10^{6}$ & $(0.6-1.8)\times 10^{6}$ & $1.8\times 10^{6}$ \\
Alfv\'{e}nic Velocity $v_A$ [cm/s]    & $4.4\times 10^{7}$ & $1.7\times 10^{7}$ & $6.9\times 10^{6}$ & $(0.3-4.8)\times 10^{6}$ & $1.0\times 10^{6}$ & $8.7\times 10^{5}$ & $6.5\times 10^{5}$ \\
\makecell{\textcolor{black}{Magnetosonic Velocity} \\ \textcolor{black}{$v_{ms}$ [cm/s]}} & \textcolor{black}{$4.9\times 10^{7}$} & \textcolor{black}{$1.8\times 10^{7}$} & \textcolor{black}{$7.1\times 10^{6}$} & \textcolor{black}{$(0.6-4.9)\times 10^{6}$} & \textcolor{black}{$2.1\times 10^{6}$} & \textcolor{black}{$2.0\times 10^{6}$} & \textcolor{black}{$1.9\times 10^{6}$} \\
\makecell{Ion Thermal Velocity \\ $v_{th,i}$ [cm/s]} & $1.4\times 10^{7}$ & $4.0\times 10^{6}$ & $1.0\times 10^{6}$ & {$(0.3-1.0)\times 10^{6}$} & $1.0\times 10^{6}$ & $(0.3-1.0)\times 10^{6}$ & $1.0\times 10^{6}$ \\
\makecell{Collisional Mean-Free-Path \\ $\lambda_{mfp}$ [cm]} & $8.8 \times 10^{-1}$ & $ 2.5 \times 10^{14}$ & $2.7\times 10^{16}$ & $(0.1-3.1)\times 10^{9}$ & $1.2\times 10^{8}$ & $(0.2-5.5)\times 10^{8}$ & $6.6\times 10^{8}$ \\
\makecell{Ion Larmor Radius \\ $r_{L,i}$ [cm]} & \textcolor{black}{$7.8\times 10^{-2}$} & $2.0\times 10^{6}$ & $3.7\times 10^{8}$ & {$ (0.3 - 7.8) \times 10^{5}$} & $7.0 \times 10^{4}$ & $ 1.3\times 10^{5}$ & $2.1\times 10^{5}$ \\ \midrule
$\lambda_{mfp}/r_{L,i}$ & \textcolor{black}{$12.2$} & $1.2\times 10^{8}$ & $7.4\times 10^{7}$ & $(1.5-700.0)\times 10^{2}$ & $1.7 \times 10^{3}$ & $(0.2-4.2)\times 10^{3}$ & $3.2\times 10^{3}$ \\
Plasma Thermal Beta $\beta_t$ & $3.0\times 10^{-1}$ & $1.3\times 10^{-1}$ & $8.1\times 10^{-2}$ & $0.02-50.0$ & $3.6$ & $(0.5-5.0)$ & $9.0$ \\
Plasma Dynamic Beta $\beta_d$ & $2.4\times 10^{1}$ & $1.7\times 10^{1}$ & $5.2\times 10^{1}$ & $0.3-240.0$ & $1.9\times 10^{2}$ & $6.6\times 10^{1}$ & $1.7\times 10^{2}$ \\
Mach Number $M$ & $6.8$ & $8.8$ & $2.0\times 10^{1}$ & $1.1-5.3$ & $5.6$ & $(2.8-8.9)$ & $3.4$ \\
Alfv\'{e}nic Mach Number $M_A$    & $3.4$ & $3.0$ & $5.1$ & $0.4-11.0$ & $9.7$ & $5.7$ & $9.2$ \\
\makecell{\textcolor{black}{Magnetosonic Mach} \\ \textcolor{black}{Number $M_{ms}$}}& \textcolor{black}{$3.1$} & \textcolor{black}{$2.8$} & \textcolor{black}{$4.9$} & \textcolor{black}{$0.4-4.8$}  & \textcolor{black}{$4.8$} & \textcolor{black}{$2.5-4.7$} & \textcolor{black}{$3.2$} \\
Reynolds Number $R_e$ & $1.5\times 10^{2}$ & $1.2\times 10^{2}$ & $ 1.0 \times 10^{2}$ & $(2.4-630.0)\times 10^{13}$ & $4.5\times 10^{15}$ & $(0.6-6.0)\times 10^{15}$ & $5.4\times 10^{14}$ \\
\makecell{Magnetic Reynolds \\ Number $R_{e_M}$} & $1.8\times 10^{5}$ & $2.8\times 10^{10}$ & $1.0\times 10^{11}$ & $(2.8-110.0) \times 10^{19}$ & $4.1\times 10^{21}$ & $(0.1-1.9)\times 10^{21}$ & $2.3\times 10^{21}$ \\
Peclet Number $P_e$ & $3.5$ & $3.0$ & $2.4$ & $(5.6-1500)\times 10^{11}$  &$1.0\times 10^{14}$ & $(0.1-1.4)\times 10^{14}$ & $1.3\times 10^{13}$ \\ \bottomrule
\end{tabular}%
\caption{\linespread{1.0}\selectfont{} \textbf{Comparison between the parameters of the shocks produced in our experiment}, with the ones of the Earth solar wind interacting at two locations (at the Earth's bow shock \cite{slavin1981solar,bale2003density,liebert2018statistical,stasiewicz2020quasi,balogh2013physics} and at the termination shock outside the solar system \cite{richardson2008cool, burlaga2008magnetic}) and of the mixed morphology SNRs interacting with a dense molecular cloud (i.e. W28 \cite{Velaz2002, hoffman2005oh,brogan2007oh,li2010gamma,  abdo2010fermi, vaupre2014, Okon2018, phan2020}, Kes. 78 \cite{miceli2017xmm,boumis2009discovery,koralesky1998shock,lockett1999oh}, W44 \cite{frail1998oh,claussen1997polarization}, and IC443 \cite{greco2018discovery,cesarsky1999isocam,reach2019supernova}).
Ion species are dominated by protons. For the SNRs with velocity less than $400$ km/s, ion temperatures are assumed to be equal to electron temperatures \cite{ghavamian2006physical}, which refer to immediate post-shock values. Since the typical radiative cooling length is much larger than the Larmor radius for cosmic rays \cite{kaufman1996far}, the temperatures used here should not be affected by it.
Numbers in bold are the primary ones, either measured in our experiment or inferred from the cited publications for the natural plasmas. 
Numbers in light are derived from the averaged primary numbers. 
The thermal (resp. dynamic) beta parameter is the ratio of the plasma thermal (resp. ram) pressure over the magnetic pressure. 
The Mach number is the ratio of the flow velocity over the sound velocity, the Alfv\'{e}nic Mach Number is the ratio of the flow velocity over the Alfv\'{e}n velocity, \textcolor{black}{and the Magnetosonic Mach number is the ratio of the flow velocity over the Magnetosonic velocity}. 
}
\label{SI_tab:astro_details}
\end{sidewaystable}

\section*{FIGURE FILES}
\renewcommand{\figurename}{{\bf Figure}}
\renewcommand{\thefigure}{{\bf \arabic{figure}}}
\setcounter{figure}{0}

\begin{figure}[htp]
    \centering
    \includegraphics[width=\textwidth]{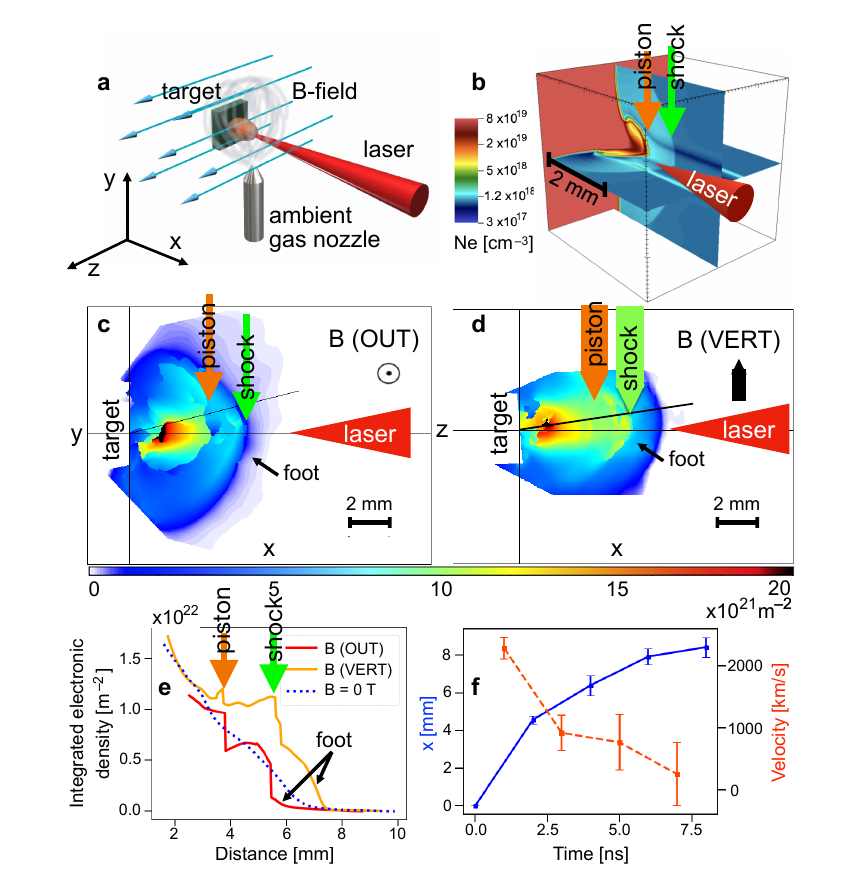}
    \caption{\textbf{Configuration and characterization of the laboratory super-critical shock.}}
\end{figure}

\begin{figure}[htp]
    \centering
    \includegraphics[width=\textwidth]{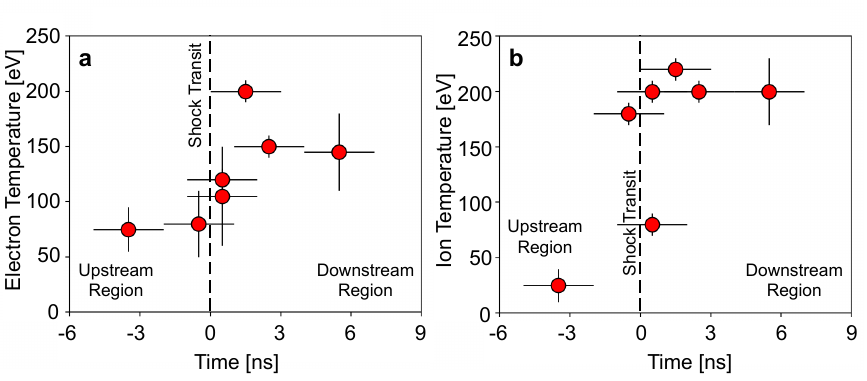}
    \caption{\textbf{Laboratory characterization of electron and ion temperature in the shock.}}
\end{figure}

\begin{figure}[htp]
    \centering
    \includegraphics[width=\textwidth]{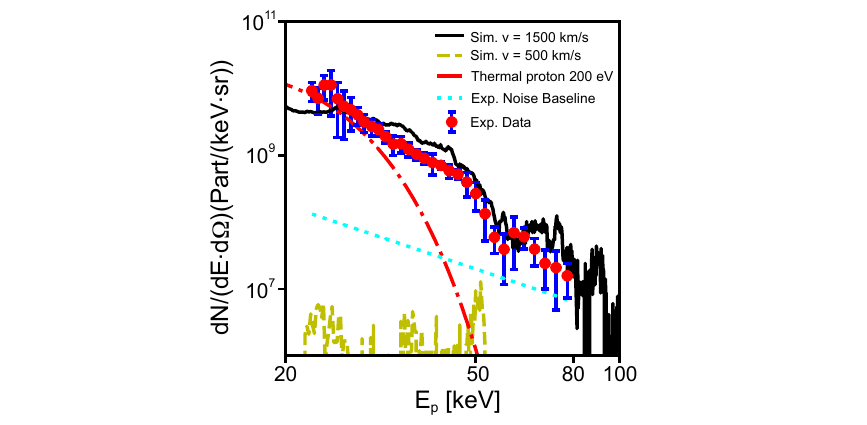}
    \caption{\textbf{Evidence for the energization of protons picked up from the \textcolor{black}{ambient medium}.}}
\end{figure}

\begin{figure}[htp]
    \centering
    \includegraphics[width=\textwidth]{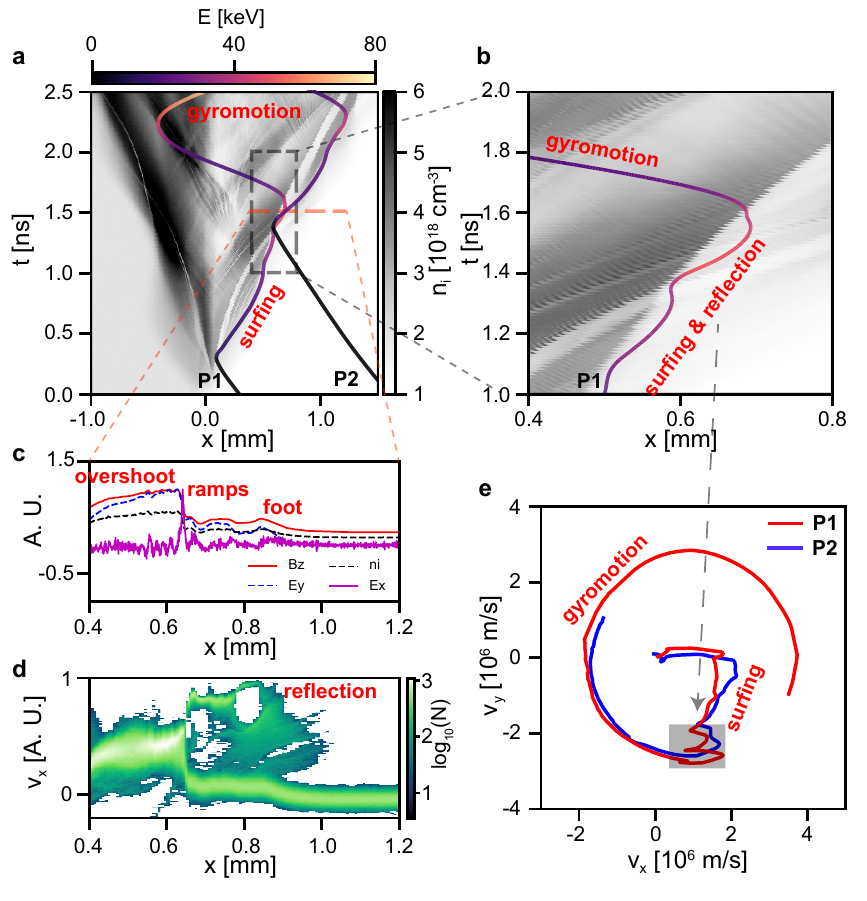}
    \caption{\textbf{Dynamics of the shock surfing proton energization from PIC simulations.} }
\end{figure}

\end{document}